\documentclass[prl,twocolumn,showpacs,superscriptaddress]{revtex4-1}

\usepackage{epstopdf}
\usepackage{amsbsy}
\usepackage{amsmath}
\usepackage{amsfonts}
\usepackage{amsthm}
\usepackage{color}
\usepackage{mathrsfs}
\usepackage{graphicx}
\usepackage{dcolumn}
\usepackage{bm}

 \usepackage{amsmath}    \usepackage[normalem]{ulem}
\usepackage{bm}
\usepackage[normalem]{ulem}
\newcommand{\ket}[1]{\vert#1\rangle}
\newcommand{\bra}[1]{\langle#1\vert}
\def\proj#1#2{\ket{#1}\bra{#2}}
\def\opone{\leavevmode\hbox{\small1\kern-3.8pt\normalsize1}}

\begin{document}

\title{Real-world two-photon interference and proof-of-principle quantum key distribution immune to detector attacks}
 
\author{A.~Rubenok}
\affiliation{Institute for Quantum Science \& Technology, University of Calgary, Canada}
\affiliation{Department of Physics \& Astronomy, University of Calgary, Canada}
\author{J.~A.~Slater}
\affiliation{Institute for Quantum Science \& Technology, University of Calgary, Canada}
\affiliation{Department of Physics \& Astronomy, University of Calgary, Canada}
\author{P.~Chan}
\affiliation{Institute for Quantum Science \& Technology, University of Calgary, Canada}
\affiliation{Department of  Electrical and Computer Engineering, University of Calgary, Canada}
\author{I. Lucio-Martinez}
\affiliation{Institute for Quantum Science \& Technology, University of Calgary, Canada}
\affiliation{Department of Physics \& Astronomy, University of Calgary, Canada}
\author{W.~Tittel}
\affiliation{Institute for Quantum Science \& Technology, University of Calgary, Canada}
\affiliation{Department of Physics \& Astronomy, University of Calgary, Canada}

\begin{abstract}
Several vulnerabilities of single photon detectors have recently been exploited to compromise the security of quantum key distribution (QKD) systems. In this letter we report the first proof-of-principle implementation of a new quantum key distribution protocol that is immune to any such attack. More precisely, we demonstrated this new approach to QKD in the laboratory over more than 80 km of spooled fiber, as well as across different locations within the city of Calgary. The robustness of our fibre-based implementation, together with the enhanced level of security offered by the protocol, confirms QKD as a realistic technology for safeguarding secrets in transmission. Furthermore, our demonstration establishes the feasibility of controlled two-photon interference in a real-world environment, and thereby removes a remaining obstacle to realizing future applications of quantum communication, such as quantum repeaters and, more generally, quantum networks.  \end{abstract}

\maketitle

Quantum key distribution (QKD) promises the distribution of cryptographic keys whose secrecy is guaranteed by fundamental laws of quantum physics\cite{Gisin2002,Scarani2009}. Starting with its invention in 1984\cite{BB84}, theoretical and experimental QKD have progressed rapidly. Information theoretic security, which ensures that secret keys can be distributed even if the eavesdropper, Eve, is only bounded by the laws of quantum physics, has been proven under various assumptions about the devices of the legitimate QKD users, Alice and Bob\cite {ShorPreskill,GLLP}. Furthermore, experimental demonstrations employing quantum states of light have meanwhile resulted in key distribution over more than 100 km distance through optical fiber\cite{Stucki2009} or air\cite{Schmitt-Manderbach2007}, QKD networks employing trusted nodes\cite{Sasaki2011}, as well as in commercially available products\cite{commercialQKD}.

However, it became rapidly clear that some of the  assumptions made in  QKD proofs were difficult to meet in real implementations, which opened side channels for eavesdropping attacks. The most prominent examples are the use of quantum states encoded into attenuated laser pulses as opposed to single photons\cite{Brassard2000}, and, more recently, various possibilities for an eavesdropper to remote-control or monitor single photon detectors\cite{Lamas2007,Zhao2008, Lydersen2010a,Lydersen2010b}.  Fortunately, both side channels can be removed using appropriately modified protocols. In the first case, randomly choosing between so-called signal or decoy states (quantum states encoded into attenuated laser pulses with different mean photon numbers)  allows one to establish a secret key strictly from information conveyed by single photons emitted by the laser\cite{Hwang2003,Wang2005,Lo2005}. (We remind the reader that an attenuated laser pulse comprising on average  $\mu$ photons contains exactly one photon with probability $P_1(\mu)=\mu e^{-\mu}$ [\citenum{Brassard2000}].) Furthermore, the recently proposed measurement-device independent (MDI) QKD protocol\cite{Lo2011} (for closely related work see  [\citenum{Braunstein2011}]) additionally ensures that controlling or monitoring detectors, regardless by what means, does not help the eavesdropper to gain information about the distributed key. Note that, while the two most prominent side channels are removed by MDI-QKD, others remain open and have to be closed by means of appropriate experimental design (see the Supplemental Material).

The MDI-QKD protocol is a clever time-reversed version of QKD based on the distribution and measurement of pairs of maximally entangled photons\cite{Bennett1992b}: In the idealized version, Alice and Bob randomly and independently prepare single photons in one out of the four qubit states $\ket{\psi}_{A,B}\in [\ket{0},\ket{1},\ket{+},\ket{-}]$, where $\ket{\pm}=2^{-1/2}(\ket{0}\pm\ket{1})$.  The photons are then sent to Charlie, who performs a Bell state measurement, i.e. projects the photons' joint state onto a maximally entangled Bell state\cite{Tittel2001}. Charlie then publicly announces the instances in which his measurement resulted in a projection onto $\ket{\psi^-}\equiv2^{-1/2}\big (\ket{0}_A\otimes\ket{1}_B-\ket{1}_A\otimes\ket{0}_B\big )$ and, for these cases, Alice and Bob publicly disclose the bases (z, spanned by $\ket{0}$ and $\ket{1}$,  or x, spanned by $\ket{\pm}$) used to prepare their photons. (They keep their choices of states secret.)  Identifying quantum states with classical bits (e.g. $\ket{0},\ket{-}\equiv~$0, and $\ket{1},\ket{+}\equiv$~1) and keeping only events in which Charlie found $\ket{\psi^-}$ and they picked the same basis, Alice and Bob now establish anti-correlated key strings. (Note that a projection of two photons onto $\ket{\psi^-}$ indicates that the two photons, if prepared in the same basis, must have been in orthogonal states.) Bob then flips all his bits, thereby converting the anti-correlated strings into correlated ones. Next, the so-called \textit{x-key} is formed out of all key bits for which Alice and Bob prepared their photons in the x-basis; its error rate is used to bound the information an eavesdropper may have acquired during photon transmission. Furthermore, Alice and Bob form the \textit{z-key} out of those bits for which both picked the z-basis.  Finally, they perform error correction and privacy amplification\cite{Gisin2002,Scarani2009} to the \textit{z-key},  which results in the secret key.

As in the entanglement-based protocol, the time-reversed version ensures that Eve cannot gain information by eavesdropping photons during transmission  or by modifying the device that generates entanglement -- either the source of photon pairs or the projective two-photon measurement, respectively -- without leaving a trace\cite{Biham1996,Inamori2002}.   Furthermore, the outstanding attribute of the MDI-QKD protocol is that it de-correlates  detection events (here indicating a successful projection onto the $\ket{\psi^-}$ Bell state) from the values of the $x$- and \textit{z-key} bits and hence the secret key bits. In other words, all side channels related to the detection setup, regardless whether actively attacked or passively monitored, do not help Eve  gain information about the secret key.  

Unfortunately, the described procedure is currently difficult to implement for two reasons, first of which is the lack of practical single photon sources. However, it is possible to replace the true single photons by attenuated laser pulses of varying mean photon number (i.e. signal and decoy states, as introduced above), and to establish the secret key using information only from joint measurements at Charlie's that stem from Alice and Bob both sending single photons\cite{Wang2012}. This procedure results in the same security against eavesdropping as the conceptually simpler one discussed above. The secret key rate, $S$, distilled from signal states, is then given by\cite{Lo2011}:
\begin{equation}
S \geq Q_{11}^z\big (1-h_2(e_{11}^x)\big ) - Q_{\mu\sigma}^zf h_2(e_{\mu\sigma}^z  ),
\label{eq:secret_key_rate}
\end{equation}
\noindent
where $h_2(X)$ denotes the binary entropy function evaluated on $X$, and $f$ describes the efficiency of error correction with respect to Shannon's noisy coding theorem. Furthermore, $Q_{11}^z$, $e_{11}^x$, $Q_{\mu\sigma}^z$,  and $e_{\mu\sigma}^z$ are gains ($Q$ -- the probability of a projection onto $\ket{\psi^-}$ per emitted pair of pulses) and error rates ($e$ -- the ratio of erroneous to total projections onto $\ket{\psi^-}$) in either the $x$- or $z$-basis for Alice and Bob sending single photons (denoted by subscript ``11"), or for pulses emitted by Alice and Bob with mean photon number $\mu$ and $\sigma$ (denoted by subscript ``$\mu\sigma$"), respectively. While the latter are directly accessible from experimental data, the former have to be calculated using a decoy state method~\cite{Lo2011,Wang2012} (see the Supplemental Material).

\begin{figure}
 \includegraphics[width=\columnwidth]{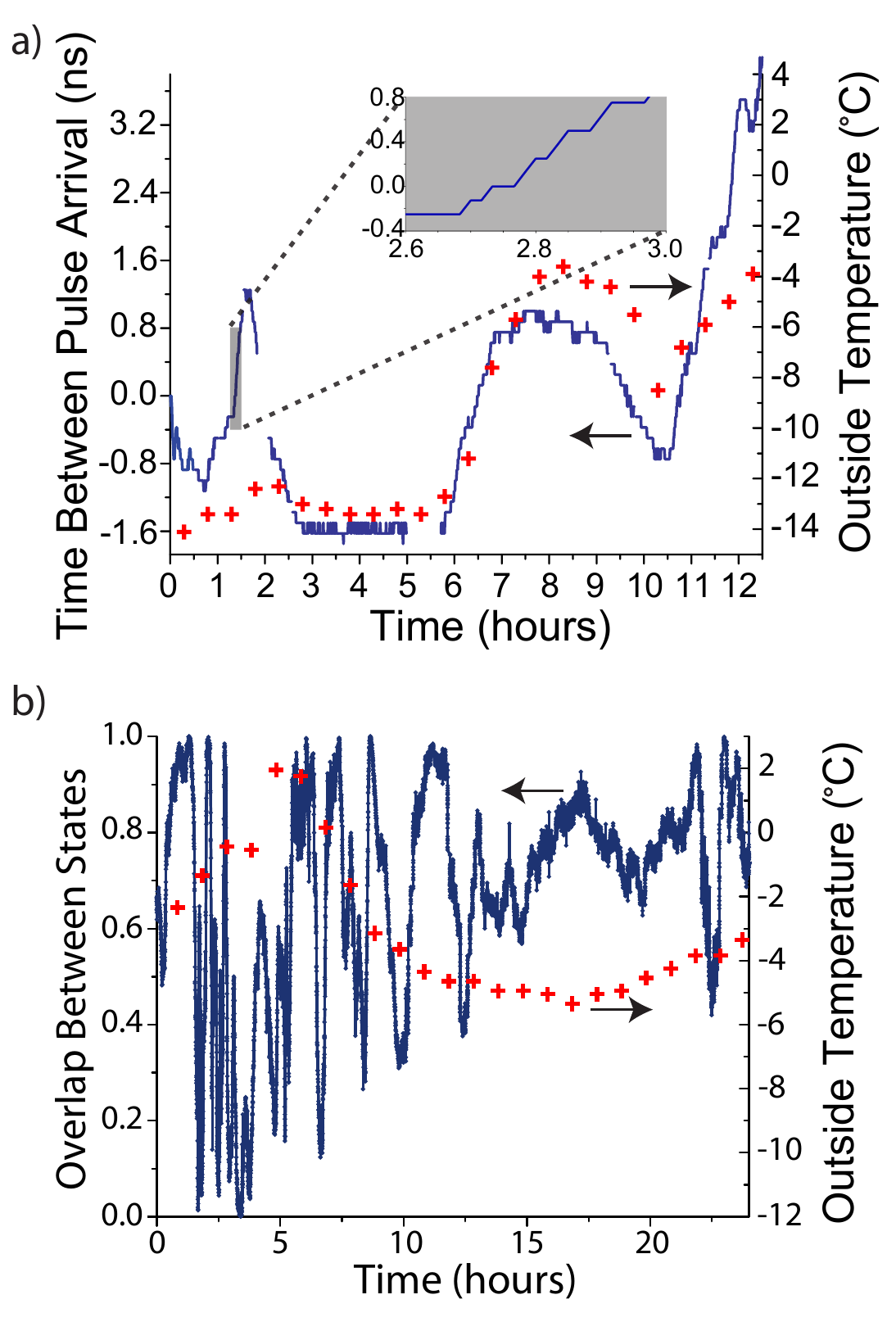}
   	\begin{center}
   \caption{\label{fig:drift} (a) Drift of differential arrival time.  Variation of arrival time difference of attenuated laser pulses emitted at Alice's and Bob's after propagation to Charlie. (b) Variation in the overlap of the polarization states of originally horizontally polarized light (emitted by Alice and Bob) after propagation to Charlie.  
Both panels include temperature data (crosses), showing correlation between variations of indistinguishability and temperature. In addition, despite local frequency locks, the  difference between the frequencies of Alice's and Bob's lasers  varied by up to 20 MHz per hour (not shown).
}
\end{center}
\end{figure}

Second, a crucial element for MDI-QKD as well as future quantum repeaters and networks is a Bell state measurement (BSM)\cite{Sangouard2011}. However, this two-photon interference measurement has not yet been demonstrated with photons that were generated by independent sources and have travelled through separate deployed fibers (i.e. fibers that feature independent changes of propagation times and polarization transformations). To implement the BSM one requires that these photons  be indistinguishable, i.e.  arrive simultaneously within their respective coherence times, with equal polarization, and feature sufficient spectral overlap. Yet, due to time-varying properties of optical fibers in a real-world environment, significant changes to photons' indistinguishability can happen in less than a minute, as depicted in Fig.~\ref{fig:drift}. Furthermore, the carrier frequencies of the signals generated at Alice's and Bob's generally vary. These instabilities make real-world Bell state measurements without stabilization by means of active feedback impossible. 

\begin{figure}
\includegraphics[angle=0,width=\columnwidth]{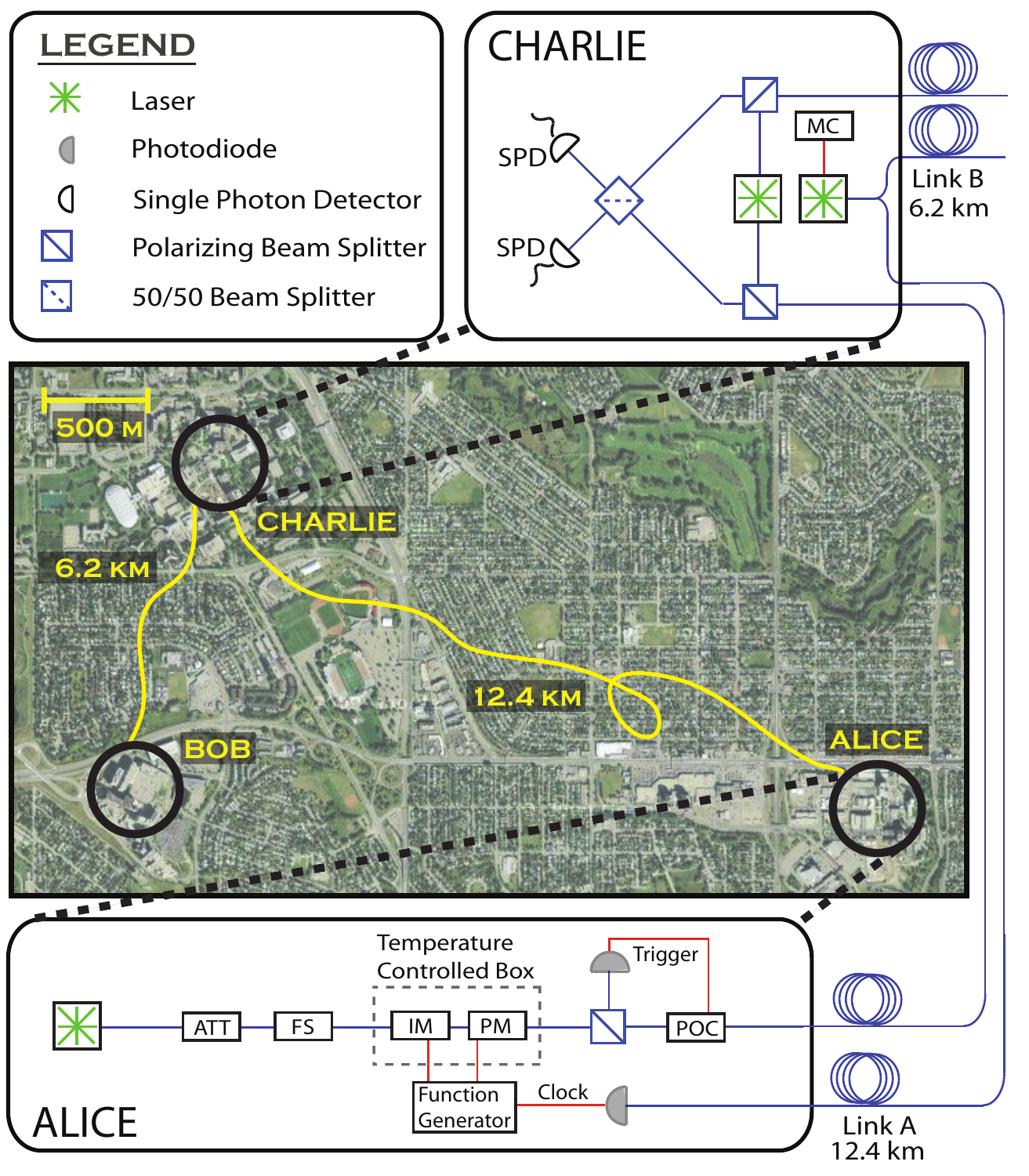}
	\begin{center}
             \caption{\label{fig:setup} Aerial view showing Alice (located at SAIT Polytechnic), Bob (located at the University of Calgary (U of C) Foothills campus) and Charlie (located at the U of C main campus). Also shown is the schematic of  the experimental setup. Optically synchronized using a master clock (MC) at Charlie's, Alice and Bob (not shown; setup identical to Alice's) generated time-bin qubits at 2 MHz rate encoded into Fourier-limited attenuated laser pulses using highly stable continuous-wave lasers at 1552.910 nm wavelength, temperature-stabilized intensity and phase modulators (IM, PM), and variable attenuators (ATT). The two temporal modes defining each time-bin qubit were of 500 ps (FWHM) duration and were separated by 1.4 ns. The qubits travelled through 12.4 and 6.2 km of deployed optical fibers to Charlie, where a 50/50 beam splitter followed by two gated ($10~\mu s$ deadtime) InGaAs single photon detectors (SPD) allowed projecting the bi-partite state onto the $\ket{\psi^-}$ Bell state. (This projection occurred if the two detectors indicate detections with 1.4$\pm$0.4 ns time difference.) The MC, polarization controller (POC) and Alice's frequency shifter (FS) are used to maintain indistinguishability of the photons upon arrival at Charlie. These three feedback systems are detailed in the Supplemental Material. The individual setups for measurements using spooled fiber (arrangement (i)) are identical. } 
	\end{center}
\end{figure}

Hence, to enable MDI-QKD and pave the way for quantum repeaters and quantum networks, we developed the ability to track and stabilize photon arrival times  and polarization transformations as well as the frequency difference between Alice's and Bob's lasers during all measurements (for more information see the Supplemental Material). 
In order to ensure the indistinguishability of photons arriving at
Charlie's and to allow, for the first time, Bell state measurements in
a real-world environment, we developed and implemented three
stabilization systems (see Fig.~\ref{fig:setup}):  fully-automatic polarization stabilization, manual
adjustment of photon arrival time, and manual adjustment of laser
frequency.  Note that automating the frequency and timing
stabilization systems is straightforward, particularly if the active
control elements  are placed in Charlie's setup.

We verified that we could indeed maintain the indistinguishability of the photons by frequently measuring the visibility, $V_{HOM}$, of the so-called Hong-Ou-Mandel dip\cite{Hong1987} (a two-photon interference experiment that is closely related to a BSM). On average we found $V_{HOM}$=47$\pm$1\%, which is close to the maximum value of 50\% for attenuated laser pulses with a Poissonian photon number distribution\cite{classicalHOM}, and thereby confirm that  real-world two-photon interference is possible.

To assess the feasibility of MDI-QKD, we implemented a proof-of-principle demonstration of MDI-QKD using the decoy state protocol proposed by Wang\cite{Wang2012}. This protocol requires that Alice and Bob choose between three different mean photon numbers: two non-zero values referred to as signal and decoy as well as vacuum. We performed our experiments over four different distances (henceforth referred to as setups) comprising two different arrangements (see Fig.~\ref{fig:setup}): (i) Alice, Bob and Charlie are located within the same lab, and Alice and Bob are connected to Charlie via separate spooled fibers of various lengths and loss. (ii) Alice, Bob and Charlie are located in different locations within the city of Calgary, and Alice and Bob are connected to Charlie by deployed fibers of 12.4 and 6.2 km length, respectively. The fiber lengths and loss in each setup are listed in Table 1.

\begin{table}
\begin{center}
\begin{tabular}{|c|c|}
\begin{tabular}{c|c | c |c | c | c | c|c }
   Setup & Fiber & $\ell_{A}$   & $l_{A}$  &$\ell_{B}$ &$l_{B}$ &total length & total loss \\
  & &  [km] &   [dB]& [km] & [dB]& $\ell$ [km]& $l$ [dB]  \\
  \hline\hline
 1a& Spool & 22.85 & \phantom{0}4.6 &22.55 & \phantom{0}4.5&\phantom{0}45.40&\phantom{0}9.1  \\
   \hline
 1b& Spool & 30.98 & \phantom{0}6.8 &34.65 & \phantom{0}6.9&\phantom{0}65.63&13.7 \\
 \hline
  1c &Spool & 40.80 & \phantom{0}9.1 &40.77 & \phantom{0}9.1 &\phantom{0}81.57&18.2 \\
     \hline
 2&Deployed & 12.4\phantom{0} & \phantom{0}4.5 & \phantom{0}6.2\phantom{0} & \phantom{0}4.5&\phantom{0}18.6\phantom{0}& \phantom{0}9.0 \\
   \hline
\end{tabular}
\end{tabular}
\end{center}
\caption{Length and loss ($\ell_{A}$, $l_{A}$, $\ell_{B}$, $l_{B}$) of the individual fiber links used to connect Alice and Charlie, and Charlie and Bob, respectively, for all tested setups. The table also lists the total length $\ell$ and total loss $l=l_{A}+l_{B}$ (in dB). The last line details measurements outside the laboratory with deployed fiber.}
\label{table_results}
\end{table}

For each setup, we prepared all 4 combinations of Alice and Bob picking a state from the z-basis (i.e. $\ket{\psi}_{A,B}\in[\ket{0},\ket{1}]$, where $\ket{0}$ and $\ket{1}$ denote time-bin qubits\cite{Tittel2001} prepared in an early or late temporal mode), and all 4 combinations of picking a state from the x-basis (i.e. $\ket{\psi}_{A,B}\in[\ket{+},\ket{-}]$).  Using a detailed model of our MDI-QKD system\cite{ourModel}, we calculated the signal and decoy intensities that maximize the secret key rate produced by the decoy-state method for each setup. For our decoy intensity we generated attenuated laser pulses containing on average $\mu = \sigma = 0.05 \pm 5\%$ photons and for our signal intensities we used a mean photon number between $0.25$ and $0.5$ (the optimal value depends on loss).  For each of the four distance configurations listed in Table 1, and for each of the 16 pairs of qubit states, we performed measurements of all 9 combinations of Alice and Bob using the signal, decoy or vacuum intensity. We recorded the number of joint detections in which one detector indicated an early arriving photon (or an early noise count), and the other detector indicated a late arriving photon (or a late noise count), which, for time-bin qubits, is regarded as a projection onto the $\ket{\psi^-}$-state\cite{Tittel2001}. Depending on the observed detection rates, measurements took between 2  and 35 minutes. This data yields the gains, $Q_{\mu\sigma}^{z}$ and $Q_{\mu\sigma}^{x}$, and error rates, $e_{\mu\sigma}^{z}$ and $e_{\mu\sigma}^{x}$, a subset of which is plotted in Fig.~\ref{fig:results}a. A complete list of gains and error rates is presented in the Supplemental Material.

\begin{figure}
 \includegraphics[angle=0,width=\columnwidth]{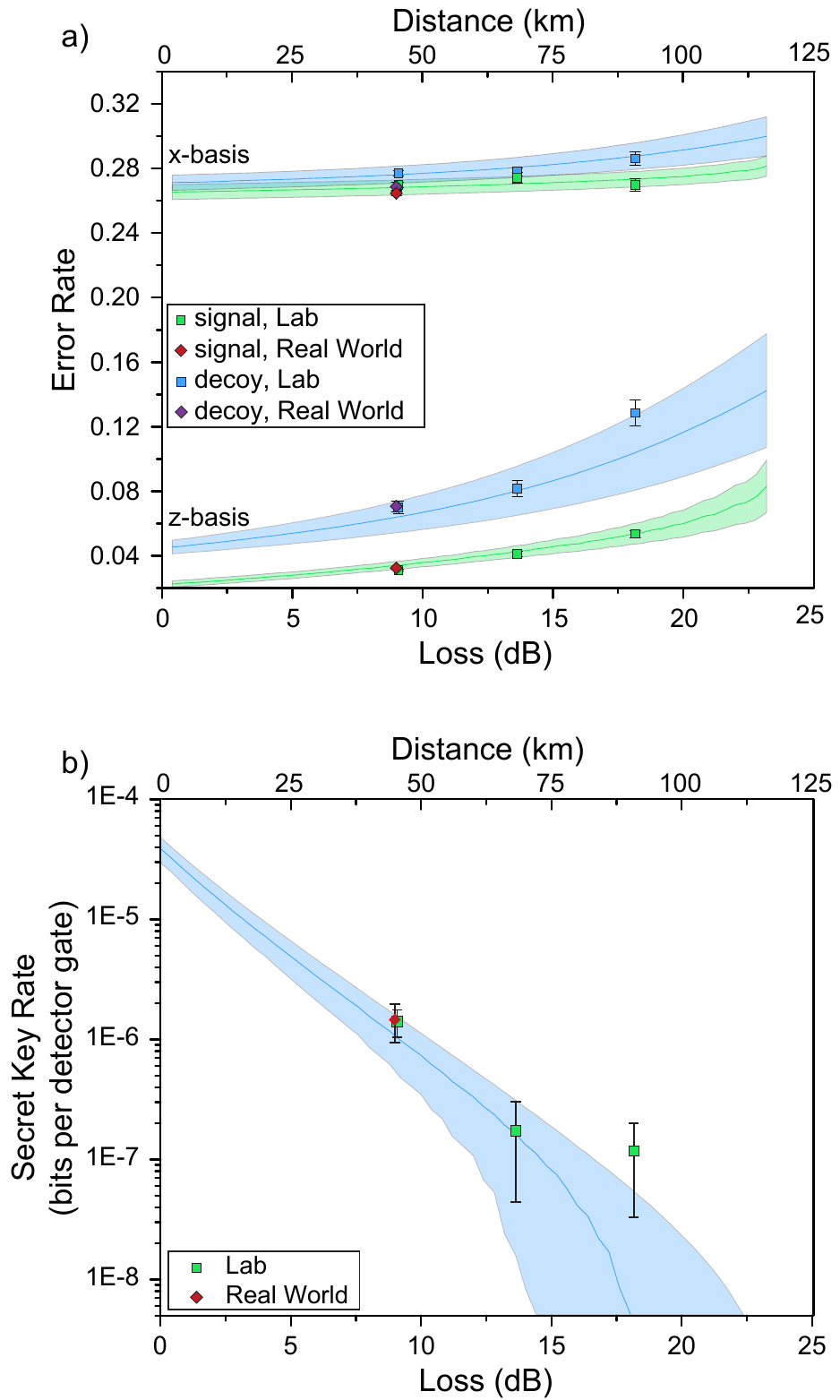}
	\begin{center}
          \caption{\label{fig:results} (a) Measured error rates $e_{\mu\sigma}^{z}$ and $e_{\mu\sigma}^{x}$ for Alice and Bob either both using signal intensity or both using decoy intensity as a function of total loss, $l=l_A+l_B$ (in dB). We note that every other combination of intensities used in the decoy-state analysis requires Alice or Bob (or both) sending vacuum, and thus the error rate is $50\%$ and not plotted. (b) Experimentally obtained and simulated secret key rates as a function of total loss, $l=l_A+l_B$ (in dB), with $l_A\cong l_B$, for optimized mean photon numbers.  Experimental secret key rates are directly calculated from measured gains and error rates using the decoy state method\cite{Wang2012} (see Supplemental Material for details).  In both panels, the secondary x-axis shows distance assuming loss of 0.2 dB/km. Diamonds depict results obtained using deployed fibers (see Fig.~\ref{fig:setup}a); all other data was obtained using fiber on spools. Uncertainties (one standard deviation)  were calculated for all measured points assuming Poissonian detection statistics. We stress that the simulated values, computed using our model\cite{ourModel}, do not stem from fits but are based on parameters that have been established through independent measurements. Monte-Carlo simulations using uncertainties in these measurements lead to predicted bands as opposed to lines (for more details see the Supplemental Material).}
	\end{center}
\end{figure}

We then computed secret key rates according to Eq. \ref{eq:secret_key_rate} after extracting $Q_{11}^z$ and $e_{11}^x$ using Wang's decoy state calculation\cite{Wang2012} and assuming an error correction efficiency $f$=1.14\cite{Sasaki2011}. As shown in Fig.~\ref{fig:results}b, all our measurements, both outside and inside the laboratory, and using up to 80 km of spooled fiber between Alice and Bob, output a positive secret key rate. Furthermore, using our model\cite{ourModel}, we estimate that our setup allows secret key distribution up to a total loss of 18$\pm$4.8 dB, which is in agreement with our QKD results. Assuming the standard loss coefficient for telecommunication fibers without splices of 0.2 dB/km, this value corresponds to a maximum  distance between Alice and Bob of 90$\pm$24 km. Note that moving from our proof-of-principle demonstration to the actual distribution of secret keys requires additional  developments, which are detailed in the Supplemental Material.

In summary, we have demonstrated that real-world quantum key distribution with practical attenuated laser pulses and immunity to detector hacking attacks is possible using current technology. Our setup contains only standard, off-the-shelf components, its development into a complete QKD system follows well-known steps\cite{Sasaki2011}, and the extension to higher key rates using state-of-the-art detectors\cite{detectors, Marsili2012} is straightforward. We also point out that MDI-QKD is well suited for key distribution over long distances, and we expect that further developments will rapidly push the separation between Alice and Bob beyond its current maximum of 250 km\cite{Stucki2009}. Finally, we remind the reader that the demonstrated possibility for Bell state measurements in a real-world environment and with truly independent photons also removes a remaining obstacle to building a quantum repeater, which promises quantum communication such as QKD over arbitrary distances.

\textbf{Note added:} We note that related experimental work has recently been reported in http://arxiv.org/abs/1207.0392 and http://arxiv.org/abs/1209.6178.


\section*{Acknowledgements}
The authors thank E. Saglamyurek, V. Kiselyov and TeraXion for discussions and technical support, the University of Calgary's Infrastructure Services for providing access to the fiber link between the University's main campus and the Foothills campus, SAIT Polytechnic for providing laboratory space, and acknowledge funding by NSERC, QuantumWorks, General Dynamics Canada, iCORE (now part of Alberta Innovates Technology Futures), CFI, AAET and the Killam Trusts.

\onecolumngrid
\section{Supplemental Material}
\section{Ensuring Indistinguishability}
In order to ensure the indistinguishability of photons arriving at
Charlie's and to allow Bell state measurements in
a real-world environment, we developed and implemented three
stabilization systems (see Fig.~2 in the main text):  fully-automatic polarization stabilization, manual
adjustment of photon arrival time, and manual adjustment of laser
frequency.  Note that automating the frequency and timing
stabilization systems is straightforward, particularly if the active
control elements  are placed in Charlie's setup.

The polarization stabilization system~\cite{Lucio2009,Bussieres2010}
employed an additional laser (at Charlie's) and two polarization
controllers (one at Alice's and one at Bob's). Every 10 s, Charlie
disabled data collection for 0.5 s and sent high intensity, vertically
polarized stabilization light to Alice and Bob.  This light was
detected by photodiodes at Alice's and Bob's, and used to trigger
their commercially available polarization controllers (POCs), which
were programmed to adjust the polarization of the stabilization light
to vertical. This implies that Alice's and Bob's attenuated laser
pulses, which were emitted horizontally polarized, both arrive
horizontally polarized at Charlie's.

To stabilize the frequency difference between Alice's and Bob's
lasers, Alice used a  frequency shifter (FS) that employed a linear
phase chirp via a serrodyne modulation signal applied to a phase
modulator. Whenever the error rate in the \textit{x-key}  increased
significantly, Charlie communicated the frequency difference after
measuring the beat frequency by mixing their unmodulated and
unattenuated laser outputs on the beam splitter. Adjustments, in the
worst case, were required every 30 minutes to maintain the difference
below 10 MHz.

To enable temporal synchronization, Charlie sent a master clock
signal  via a second set of fibers to Alice and Bob. Roughly every
minute, Charlie measured the qubit arrival-time difference using his
SPDs and  high-resolution electronics and sent this information to
Alice and Bob. They then adjusted their qubit generation times using
function generators to apply a phase shift to the recovered master
clock. This maintained the arrival-time difference under 30~ps.

\section{Decoy-State Analysis}
In MDI-QKD the secret key rate is given by 
\begin{equation}
S \geq Q_{11}^z\big (1-h_2(e_{11}^x)\big ) - Q_{\mu\sigma}^zf h_2(e_{\mu\sigma}^z  ),
\end{equation}
where $h_2(X)$ denotes the binary entropy function evaluated on $X$, and $f$ describes the efficiency of error correction with respect to Shannon's noisy coding theorem. Furthermore, $Q_{11}^z$, $e_{11}^x$, $Q_{\mu\sigma}^z$,  and $e_{\mu\sigma}^z$ are gains ($Q$ -- the probability of a projection onto $\ket{\psi^-}$ per emitted pair of pulses) and error rates ($e$ -- the ratio of erroneous to total projections onto $\ket{\psi^-}$) in either the $x$- or $z$-basis for Alice and Bob sending single photons (denoted by subscript ``11"), or for pulses emitted by Alice and Bob with mean photon number $\mu$ and $\sigma$ (denoted by subscript ``$\mu\sigma$"), respectively. While $Q_{\mu\sigma}^z$,  and $e_{\mu\sigma}^z$ are directly accessible from experimental data, $Q_{11}^z$, $e_{11}^x$ have to be bounded using a decoy state method.

We use a three-intensity decoy state method for the MDI-QKD protocol~\cite{Wang2012} that derives a lower bound for $Q_{11}^{x}$ and $Q_{11}^{z}$ and an upper bound for $e_{11}^x$, to calculate a lower bound for the secure secret key rate. We denote the signal, decoy, and vacuum intensities by $\mu_s$, $\mu_d,$ and $\mu_v$,
respectively, for Alice, and Bob (note that $\mu_v = 0$ by definition). In our implementation Alice and Bob both select the same mean
photon numbers for the three intensities and use channels of equal transmission. For compactness of
notation, we omit the $\mu$ when describing the gains and
error rates (e.g. we write $Q_{ss}^z$ to denote the gain in the z-basis
when Alice and Bob both send photons using the signal intensity).  Under
these assumptions, the lower bound on $Q_{11}^{x}$ is given by\\

\begin{equation}
Q_{11}^{x} \ge \frac{ P_1(\mu_s)P_2(\mu_s)\big(Q_{dd}^{x} -
Q_0^{x}(\mu_d)\big) - P_1(\mu_d)P_2(\mu_d)\big(Q_{ss}^{x} -
Q_0^{x}(\mu_s)\big)
}{P_1(\mu_s)P_1(\mu_d)\big(P_1(\mu_d)P_2(\mu_s)-P_1(\mu_s)P_2(\mu_d)\big)},
\label{eqn:Q11_bound}
\end{equation}

\noindent
where the various $P_i(\mu)$ denote the probabilities that a pulse with
Poissonian photon number distribution and mean $\mu$ contains exactly $i$
photons, and $Q_{0}^{z}(\mu_d)$ and $Q_{0}^{z}(\mu_s)$ are given by

\begin{eqnarray}
Q_{0}^{x}(\mu_d) & = & P_0(\mu_d)Q_{vd}^{x} + P_0(\mu_d)Q_{dv}^{x} -
P_0(\mu_d)^2Q_{vv}^{x},\\
Q_{0}^{x}(\mu_s) & = & P_0(\mu_s)Q_{vs}^{x} + P_0(\mu_s)Q_{sv}^{x} -
P_0(\mu_s)^2Q_{vv}^{x}.
\label{eqn:Q0}
\end{eqnarray}

\noindent
Similar equations are used to bound $Q_{11}^{z}$ (we replace the superscript $x$ by $z$). Finally, the error rate
$e_{11}^x$ can then be computed as

\begin{equation}
e_{11}^x \le \frac{e_{dd}^xQ_{dd}^x - P_0(\mu_d)e_{vd}^xQ_{vd}^x -
P_0(\mu_d)e_{dv}^xQ_{dv}^x +
P_0(\mu_d)^2e_{vv}^xQ_{vv}^x}{P_1(\mu_d)^2Q_{11}^x},
\label{eqn:e11_bound}
\end{equation}

\noindent
where the upper bound holds if a lower bound is used for $Q_{11}^x$.  Note
that $Q_{11}^{x,z}$, $Q_{0}^{x,z}(\mu_d)$, $Q_{0}^{x,z}(\mu_s)$ and
$e_{11}^x$ (Eqs.~\ref{eqn:Q11_bound}-\ref{eqn:e11_bound}) are uniquely
determined through measurable gains and error rates.

Our analysis in [\citenum{ourModel}] determined that lowering $\mu_d$ as much as possible maximizes secret key rate. In these experiments, we select $\mu_d = 0.05$ in order to obtain statistically significant data in a reasonable amount of time (see Suplementary Table~\ref{table_results2})

\begin{table*}[t]
\caption{List of experimentally obtained error rates, $e_{\mu\sigma}^{x,z}$, and gains, $Q_{\mu\sigma}^{x,z}$, used to calculate the secret key rate in four different configurations. For each configuration we show the mean photon numbers for the signal and decoy states, $\mu_s$ and $\mu_d$, employed by Alice and Bob. The vacuum state corresponds to a mean photon number of $\mu_v=0$. We remind the reader that we omit the $\mu$ when writing the gains and error rates, writing only the subscript denoting the signal ($s$), decoy ($d$), or vacuum ($v$) state.  We also indicate the lengths of fiber connecting Alice and Charlie ($\ell_{A}$), Bob and Charlie ($\ell_{B}$) and the total transmission loss ($l$). Finally, the computed secret key rate ($S$) is shown in bits per detector gate. Additionally, we measured $Q_{vv}^{x,z} = (7.1\pm 0.30)\times10^{-10}$ and $e_{vv}^{x,z} = 0.49 \pm 0.021$, which is applied to all distances.}

\begin{center}
\begin{tabular}{c}
\begin{tabular}{|c | c || c | c | c | c|| c | c | c | c|}
Fiber & Spool &\multicolumn{4}{c||}{Z-basis} & \multicolumn{4}{c|}{X-basis}\\
\hline
         $\ell_{A}$ & 22.85 km & $Q_{ss}^ z$ & $1.028(3) \times10 ^{-4} $ & $e_{ss}^z$ & 0.0311(4)  &  $Q_{ss}^x$ &   $1.95(1) \times 10^{-4}$ &  $e_{ss}^x$ &  0.270(2)\\
         \hline
         $\ell_{B}$ &  22.55 km  &  $Q_{sv}^z$ & $2.98(5)\times 10^{-6}$  & $e_{sv}^z$  &0.49(1) &   $Q_{sv}^x$ & $5.68(2) \times 10^{-5}$   & $e_{sv}^x$ & 0.494(2)\\
         \hline
         Total loss $l$  &  9.1 dB  & $Q_{vs}^z$  & $1.78(4)\times 10^{-6}$ & $e_{vs}^z$ &  0.47(1)   & $Q_{vs}^x$   & $5.77(2) \times 10^{-5}$  & $e_{vs}^x$  & 0.507(2)\\
         \hline
         $\mu_s$ & 0.396(4) & $Q_{dd}^z$ & $1.89(3)\times 10^{-6} $ &  $e_{dd}^z$ & 0.070(4)  & $Q_{dd}^x$ & $3.40(1) \times 10^{-6}$  & $e_{dd}^x$  & 0.277(2)\\
\hline         
         $\mu_d$ & 0.050(1) & $Q_{dv}^z$  & $1.05(6)\times 10^{-7}$   & $e_{dv}^z$ & 0.47(3)  & $Q_{dv}^x$ & $8.76(8)\times 10^{-7}$  & $e_{dv}^x$  & 0.511(5) \\
         \hline
          $S$ & $1.4(4)\times 10^{-6}$ & $Q_{vd}^z$  & $9.24(5)\times 10^{-8}$ & $e_{vd}^z$  & 0.48(3)  & $Q_{vd}^x$ & $8.59(9)\times 10^{-7}$ & $e_{vd}^x$ & 0.503(5)\\
\hline
\multicolumn{10}{c}{}	\\
Fiber & Spool  &  \multicolumn{4}{c||}{Z-basis} & \multicolumn{4}{c|}{X-basis}\\
\hline
         $\ell_{A}$ & 30.98 km & $Q_{ss}^ z$  & $1.67(1) \times10^{-5}$  & $e_{ss}^z$  & $0.041 (2) $ &  $Q_{ss}^x$ &  $3.57(3) \times 10^{-5}$  & $e_{ss}^x$ & 0.274(3)\\
         \hline
         $\ell_{B}$ & 34.65 km  & $Q_{sv}^z$ &  $6.7(2) \times 10^{-7}$  & $e_{sv}^z$  &  0.51(2) &   $Q_{sv}^x$ &  $9.62(9) \times 10^{-6}$  &  $e_{sv}^x$ & 0.498(4) \\
         \hline
         Total loss $l$ & 13.7 dB & $Q_{vs}^z$  & $4.4(2) \times  10^{-7}$  &  $e_{vs}^z$ &  0.48(2)  & $Q_{vs}^x$   & $9.32(7) \times 10^{-6}$  & $e_{vs}^x$  & 0.499(4) \\
\hline
         $\mu_s$ & 0.279(6) & $Q_{dd}^z$ &  $6.0(1) \times 10^{-7}$ & $e_{dd}^z$ &  0.082(5)  & $Q_{dd}^x$ &  $1.192(7)\times 10^{-6}$  & $e_{dd}^x$  & 0.278(2)\\
         \hline
         $\mu_d$ & 0.050(1) & $Q_{dv}^z$  &  $4.7(4) \times 10^{-8}$  & $e_{dv}^z$ &  0.47(4)  & $Q_{dv}^x$ & $3.08(7) \times 10^{-7}$ & $e_{dv}^x$  & 0.50(1) \\
         \hline
          $S$  &  $1.7(1.3)\times 10^{-7}$ & $Q_{vd}^z$  & $4.0(4) \times 10^{-8}$ & $e_{vd}^z$  & 0.41(4)  & $Q_{vd}^x$ &  $3.03(7)\times 10^{-7}$  &  $e_{vd}^x$ & 0.50(1)\\
\hline 
\multicolumn{10}{c}{}	\\
Fiber & Spool  & \multicolumn{4}{c||}{Z-basis} & \multicolumn{4}{c|}{X-basis}\\
\hline
         $\ell_{A}$ & 40.80 km  & $Q_{ss}^ z$  & $5.57(6) \times 10^{-6}$  & $e_{ss}^z$  &  0.053(2) &  $Q_{ss}^x$ &  $9.87(9) \times 10^{-6}$  & $e_{ss}^x$ & 0.270(4)   \\
         \hline
         $\ell_{B}$  &  40.77 km  & $Q_{sv}^z$ &  $2.15(9)\times 10^{-7}$  & $e_{sv}^z$  & 0.51(2)  &   $Q_{sv}^x$ &   $2.50(3) \times 10^{-6}$  &  $e_{sv}^x$ & 0.505(7)  \\
         \hline
         Total loss  $l$ & 18.2 dB & $Q_{vs}^z$  &  $1.88(8) \times 10^{-7}$ &  $e_{vs}^z$ &  0.49(2)  & $Q_{vs}^x$   &  $2.95(4)\times 10^{-6}$  & $e_{vs}^x$  & 0.501(6) \\
         \hline
         $\mu_s$ & 0.251(6) & $Q_{dd}^z$ &  $2.66(6)\times 10^{-7}$  & $e_{dd}^z$ &  0.129(8)  & $Q_{dd}^x$ &  $4.49(4) \times 10^{-7}$ & $e_{dd}^x$  &  0.286(4)  \\
         	\hline
          $\mu_d$ & 0.050(1) & $Q_{dv}^z$  &  $2.8(2)\times 10^{-8}$  & $e_{dv}^z$ &   0.52(4) & $Q_{dv}^x$ &  $1.25(4)\times 10^{-7}$  & $e_{dv}^x$  &  0.51(1)  \\
         \hline
          $S$   & $1.2(8)\times10^{-7}$ & $Q_{vd}^z$  & $2.2(2) \times 10^{-8}$  & $e_{vd}^z$  & 0.45(4) & $Q_{vd}^x$&  $1.22(3)\times 10^{-7}$  &  $e_{vd}^x$ &  0.51(1)\\
\hline
\multicolumn{10}{c}{}	\\
 
Fiber & Deployed & \multicolumn{4}{c||}{Z-basis} & \multicolumn{4}{c|}{X-basis}\\
\hline
         $\ell_{A}$ & 12.4 km & $Q_{ss}^ z$  & $1.042(3) \times 10^{-4}$ & $e_{ss}^z$  &  0.0323(6)  &  $Q_{ss}^x$ &  $2.020(8) \times 10^{-4}$  & $e_{ss}^x$ & 0.265(2)  \\
         \hline
         $\ell_{B}$ & 6.2 km & $Q_{sv}^z$ &   $2.96(6) \times 10^{-6}$& $e_{sv}^z$  &  0.50(1)  &   $Q_{sv}^x$ &   $5.63(2) \times 10^{-5}$  &  $e_{sv}^x$ & 0.492(2) \\
         \hline
         Total loss $l$  & 9.0 dB  & $Q_{vs}^z$  &  $1.87(4) \times 10^{-6}$ &$e_{vs}^z$ &  0.52(1) & $Q_{vs}^x$   &  $5.10(2) \times 10^{-5}$  & $e_{vd}^x$  &  0.512(2) \\
         \hline
         $\mu_s$ & 0.402(2) & $Q_{dd}^z$ & $1.82(2) \times 10^{-6}$  & $e_{dd}^z$ &  0.071(3)   & $Q_{dd}^x$ &  $3.35(2) \times 10^{-6}$  & $e_{dd}^x$  & 0.269(3)\\
\hline
          $\mu_d$ & 0.050(1) & $Q_{dv}^z$  &  $1.15(6) \times 10^{-7}$ & $e_{dv}^z$ &  0.53(3)  & $Q_{dv}^x$ &  $8.5(1) \times 10^{-7} $ & $e_{dv}^x$  & 0.502(6)\\
         \hline
            $S$  & $1.5(5)\times 10^{-6}$  & $Q_{vd}^z$  & $8.4(5) \times 10^{-8}$ & $e_{vd}^z$  &   0.49(4)  & $Q_{vd}^x$ &  $ 8.5(1)\times 10^{-7} $ & $e_{vd}^x$ &  0.501(6) \\
\hline
\end{tabular}
\end{tabular}
\end{center}
\label{table_results2}
\end{table*}

\section{Secure key distribution using MDI-QKD}\label{SIsec:assumptions}
 
In this section we describe the assumptions underpinning secure key distribution in MDI-QKD as well as further technological and theoretical developments required for our current proof-of-principle demonstration to meet this goal.  
We note that any QKD system used to distribute secret key must be vetted against attacks arising from imperfections in its implementation\footnote{A notable exception is fully device independent QKD (DI-QKD)~\cite{Masanes2011}, which, however, is currently impossible to realize  due to the need for a loophole free violation of a Bell inequality.}. Protection against such attacks requires the development of hardware that strives to be as ideal as possible, in conjunction with the development of security proofs that are able to take into account those imperfections that inevitably remain in any realistic implementation. (Such proofs would bound the information leaked to an eavesdropper, which, in turn, allows removing it by means of privacy amplification). Even for the heavily studied prepare-and-measure BB84 protocol, this is an area of ongoing research~\cite{Woodhead2012}, and more needs to be done for the new MDI-QKD protocol. Yet, MDI-QKD constitutes a very important development in this context as it eliminates all potential attack strategies related to imperfections in the measurement apparatus, including arbitrary measurement-basis misalignment errors as well as detector attacks that have recently been shown to provide the eavesdropper full information about the key without leaving a trace~\cite{Lamas2007,Zhao2008, Lydersen2010a,Lydersen2010b}.  Remaining assumptions and required developments are:

\begin{enumerate}

\item \textbf{Quantum mechanics is correct and complete.} This assumption is generally believed to be true.

\item \textbf{Alice's and Bob's laboratories are private.} This assumption entails that no undesired signals, e.g. RF electromagnetic radiation, escape from Alice's and Bob's apparatus when working in normal conditions.  Information gain through such passive observation can be avoided using appropriate shielding, which, as is standard in academic QKD implementations, we have not spent any particular effort on. Furthermore, the assumption implies that Eve cannot actively obtain information about the experimental settings, e.g. by sending a probe, such as light, into the laboratories using the fiber that connects Alice or Bob, respectively, with the outside world, and analyzing the back reflection. This is often referred to as a Trojan horse attack~\cite{Gisin2002, Scarani2009}. And finally, Eve cannot actively influence Alice's or Bob's devices to modify their functioning. Protection against active attacks requires that the laboratories are isolated from signals sent by Eve, e.g. using optical isolators or attenuators. No such countermeasures were realized in our proof-of-principle demonstration. However, their implementation is straightforward, at least in what concerns attenuators and isolators~\cite{Sasaki2011}. We emphasize that there is no need to protect Charlie's laboratory; the MDI-QKD protocol ensures that it can even be run by the eavesdropper.

\item \textbf{Alice and Bob send phase-randomized attenuated pulses of light produced by a laser operated well above threshold.} This ensures that the generated light pulses are correctly described by the density matrix $\rho=\sum_{n}P_n(\mu)\proj{n}{n}$, where $P_n(\mu)=\frac{e^{-\mu}\mu^n}{n!}$ is the Poisson distribution with mean photon number $\mu$, and $\proj{n}{n}$ denotes the density matrix of an $n$-photon Fock state. This condition is easily met by generating every light pulse using a laser diode triggered by a short electrical pulse. However, as we carve qubits out of a laser beam with large coherence time using an intensity modulator, it is not fulfilled in our setup (more precisely, subsequent pulses are coherent). Yet, we point out that the solution to our problem is well understood and has been implemented before~\cite{Zhao2007}:  it simply requires adding a phase modulator that randomizes the global  phase of each qubit. 

\item \textbf{The mean values of photons per pulse, as well as the encoded states are chosen randomly.} No random choices have been implemented in our current proof-of-principle demonstration. Instead, we sent pulses with the same mean photon number and encoded the same qubit state during several minutes before changing the state or mean number. However, operating the phase and amplitude modulators that generate qubit states using adequate drivers connected to quantum random number generators is well understood~\cite{Sasaki2011}, and meeting the requirement of random modulation is straightforward, though time consuming.

\item \textbf{Alice and Bob generate qubits in states that are sufficiently close to those that form two maximally conjugate bases.} These states were denoted in the main text as $\ket{0}$, $\ket{1}$, $\ket{+}\equiv\frac{1}{\sqrt{2}}(\ket{0}+\ket{1})$ and $\ket{-}\equiv\frac{1}{\sqrt{2}}(\ket{0}-\ket{1})$, respectively. This assumption may currently not be satisfied (see~\cite{ourModel} for a detailed description of our experimental imperfections). For instance, considering states in different bases (for which the overlap should be 0.5), we find an average deviation of 0.074, and for different states in the same basis (for which we expect an overlap of zero), the average deviation is 0.013.  According to the analyses in~\cite{Wang2012,Tamaki2012} these overlaps, together with the current detector performance, are insufficient to securely distribute key. However, we point out that both proofs lead to very conservative bounds. For instance, the proof in~\cite{Wang2012} requires a state generation procedure that artificially increases error rates and applies non-tight bounds, and hence underestimates secure key rates. We believe that future investigations will rapidly improve proof techniques and yield higher secret key rates (and result in secret key in cases in which current proofs predict no secret key). Furthermore, we note that straightforward technological improvements allow reducing the maximum deviation from the ideal overlap values to around 1 part in 1000. For instance, this can be accomplished by reducing ringing in our pulse generation by a factor of 5, and using commercially-available, state-of-the-art intensity modulators that allow suppressing the background by an additional 10-20 dB~\cite{EOspace}. In addition, using state-of-the-art detectors with 93\% quantum efficiency and 1kHz noise~\cite{Marsili2012} leads, according to simulation results with a theoretical model of MDI-QKD that we presented in~\cite{ourModel}, to secret key rates similar to or above the ones reported in the main document, even using the conservative  approach in~\cite{Wang2012}.

\item \textbf{Sufficiently weak correlations between qubit states and all degrees of freedom not used to encode the qubit.} In principle, the various states generated by Alice and Bob could have differences in other degrees of freedom (i.e. polarization, spectral, spatial, or temporal modes), which could open a security loophole~\cite{Nauerth09} if not properly quantified and taken into account during privacy amplification. However, for MDI-QKD, the link between correlations with unobserved degrees of freedom and Eve's information gain is not yet clear. In particular, correlations are likely to degrade the visibility of the BSM, thus creating  observable errors. The upper bound on Eve's information gain, possibly zero, can only be assessed using plausible arguments based on the actual implementation of the setup supplemented by careful measurements. For instance, in our implementation, the use of a single laser to generate all qubits states and of a single-mode fiber to transmit qubits from Alice, or Bob, to Charlie, respectively, makes it highly unlikely  that correlation between states and photon spectra or spatial modes exist. Furthermore, careful programming of the function generator that generates all states through interaction with the same intensity modulator makes it very plausible that no temporal distinguishability is observable in our experiment. And finally, the polarization beam splitter at the exit of Alice's and Bob's laboratories ensures equal polarization of all time-bin qubit states.

\item \textbf{Appropriate classical post-processing of the sifted key, i.e. error correction and privacy amplification}. Note that while we have not implemented error correction, we have used a realistic estimation of the error correction efficiency~\cite{Sasaki2011} to determine the potential secret key rate of our system. Furthermore, we did not consider finite key size effects in our proof-of-principle demonstration (in other words, we assumed that we could run our QKD devices during an infinitely long time and produce an infinite amount of measured data), which, in the case of MDI-QKD, have so far only been investigated using an overly conservative approach~\cite{Song2012}. 

\item \textbf{A short secret authentication key exists before starting QKD.} This key is used to authenticate the classical communication channel during error correction and privacy amplification. As we did not implement any of these post-processing steps, we did not need any pre-established secret key. In an actual implementation, this  step can, for instance, be accomplished during a personal meeting between Alice and Bob. 

\end{enumerate}

We recall that some of the above topics are currently not as thoroughly studied for MDI-QKD as for prepare-and-measure QKD. However, the ability to close all side channels in measurement devices represents a significant step forward in closing the gap between theoretical security proofs and experimentally viable implementations. In particular, it has, for the first time, allowed for the development of security proofs in QKD that take arbitrary state generation and measurement errors into account, even though the efficiency of the current approaches can certainly be increased\footnote{In comparison, the only security proof for BB84 QKD dealing with arbitrary state generation errors at the source and arbitrary misalignment of the measurement bases is limited to individual attacks but does not apply to  more powerful coherent attacks~\cite{Woodhead2012}.}. In addition, for actual key distribution, our experimental implementation has to be improved along the lines discussed above. 
We leave these interesting and important topics for future investigations and emphasize that our work has focused on previously undemonstrated requirements for MDI-QKD, such as the Bell state measurement over deployed fiber, on improving the understanding of the capabilities and current limitations of our setup (including optimization and efficiency calculations of a decoy state analysis; for more information see~\cite{ourModel}) and on experimental demonstrations of the protocol over various distances as well as over deployed, real-world optical fiber.\\

\newpage
\section{Discussion of  error rates $e_{\mu\sigma}^{x,z}$}

Let us briefly discuss the ideal case in which the quantum states encoded into attenuated laser pulses, as well as the projection measurements, are perfect. To gain some  insight into how the difference in the error rates, $e_{\mu\sigma}^{x,z}$, arises\footnote{Note when two superscripts, each one denoting a different basis, are present on variables, (e.g. $e_{\mu\sigma}^{x,z}$, as above, or $Q_{\mu\sigma}^{x,z}$), this is a shorthand for, e.g. $e^z_{\mu\sigma}$ and $e^x_{\mu\sigma}$ -- that  is, the statement is valid for both the z- and x-bases. Note that variables may take different values for each basis, e.g. $e_{\mu\sigma}^{z} \neq e_{\mu\sigma}^{x}$.  When this notation is used within an equation such as Eq.~\ref{eqn:photon_state}, then the equation may be written for either the z- or x-basis.}, we consider only the most likely case that can cause the detection pattern associated with a projection onto $\ket{\psi^-}$ (this projection occurs if the two detectors indicate detections with 1.4$\pm$0.4 ns time difference). Specifically, we consider only the case in which two photons arrive at the beam splitter.  Note that these photons can either come from the same person, or from different persons.

\begin{itemize}
\item z-basis: Assuming that Alice and Bob both prepare states in the z-basis, only photons prepared in orthogonal states can cause a projection onto $\ket{\psi^-}$. This implies that one photon has to come from Alice, and the other one from Bob (if generated by the same person, both photons would be in the same state). Hence, taking into account Bob's bit flip, Alice and Bob always establish identical bits, i.e. $e_{\mu\sigma}^z\text{(ideal)}=0$.

\item x-basis: Assuming that both Alice and Bob  prepare states in the x-basis, it is no longer true that only photons prepared in orthogonal states and by different persons can cause a projection onto $\ket{\psi^-}$. Indeed, if the two photons have been prepared by the same person, it is possible to observe the detection pattern associated with a projection onto $\ket{\psi^-}$. In this case, given that all detected photons have been prepared by either one or the other person, the detection does not indicate any correlation between the states prepared by Alice and Bob. In turn, this leads to uncorrelated key bits. Thus, $e_{\mu\sigma}^x\text{(ideal)}$ is determined by the probability that one photon arrived from each person relative to the probability that two photons arrived from the same person.  A detailed analysis for attenuated laser pulses with Poissonian photon number distribution, assuming an equal probability of photons arriving from either party, yields $e_{\mu\sigma}^x\text{(ideal)}$ = 1/4.
\end{itemize}

\end{document}